# Reconstructing short-lived particles using hypergraph representation learning

Callum Birch-Sykes, Brian Le, Yvonne Peters, Ethan Simpson,* and Zihan Zhang†
*Department of Physics and Astronomy, University of Manchester, Manchester M13 9PL, United Kingdom*

In collider experiments, the kinematic reconstruction of heavy, short-lived particles is vital for precision tests of the Standard Model and in searches for physics beyond it. Performing kinematic reconstruction in collider events with many final-state jets, such as the all-hadronic decay of top-antitop quark pairs, is challenging. We present *Hypergraph for Particle Event Reconstruction* (HyPER), a novel architecture based on graph neural networks that uses hypergraph representation learning to build more powerful and efficient representations of collider events. HyPER is used to reconstruct parent particles from sets of final-state objects. Trained and tested on simulation, the HyPER model is shown to perform favorably when compared to existing state-of-the-art reconstruction techniques, while demonstrating superior parameter efficiency. The novel hypergraph approach allows the method to be applied to particle reconstruction in a multitude of different physics processes.

## I. INTRODUCTION

The Large Hadron Collider [1] at CERN produces $4 \times 10^7$ high-energy proton-proton collisions per second to recreate the conditions of the early Universe. These collisions frequently produce heavy, short-lived particles that promptly decay to relatively stable final states. Accurate reconstruction of these short-lived particles is critical for measuring their properties, and for defining observables which are sensitive to new physics. Further, the development of accurate and flexible methods for reconstructing short-lived particles in rare processes—like the production of top-antitop quark pairs with a Higgs boson, $t\bar{t}H$—is a vital step in studying such rare topologies and scrutinizing the Standard Model (SM).

The kinematics of short-lived particles can only be inferred by considering the recorded kinematics of their decay products. The aim of the reconstruction process is to identify which combination of decay products corresponds to each parent particle. This task becomes increasingly challenging when we consider processes with many heavy resonances, such as the simultaneous production of four top quarks, $t\bar{t}t\bar{t}$. Processes like these lead to high-multiplicity final states with large combinatorics.

*Contact author: ethan.simpson@manchester.ac.uk
†Contact author: zihan.zhang@manchester.ac.uk

The most common final-state objects produced at hadron colliders are jets: collimated sprays of neutral and charged hadrons which arise as a result of quantum chromodynamics (QCD). In addition to being produced in the decays of short-lived particles, jets can also arise from partonic splittings or concurrent proton-proton interactions. This increases the "jet multiplicity" of the final state and complicates the jet-assignment reconstruction problem. The case where heavy particles decay exclusively to jets is known as the *all-hadronic* decay channel. Particle reconstruction in this channel is simplified by the lack of final-state neutrinos, which go undetected and whose kinematics can only be partially inferred through conservation of momentum in the transverse plane.

A useful process for developing and benchmarking reconstruction techniques is the all-hadronic decay of $t\bar{t}$ pairs. Top quarks decay prior to hadronization through a near-exclusive decay channel $t \to Wb$. In all-hadronic production, both $W$ bosons then decay hadronically, with the full decay given by

$$t\bar{t} \to W^+ b W^- \bar{b} \to b q_1 q_2 \bar{b} q_3 q_4, \tag{1}$$

where the flavors of the $W$ boson decay products are distinct: $q_1 \neq q_2, q_3 \neq q_4$. All six quarks in the final state will form jets, with those originating from a $b$-quark referred to as "$b$-jets." The example Feynman diagram in Fig. 1 shows the production and all-hadronic decay of a $t\bar{t}$ pair, with two additional jets produced through QCD radiation from an initial-state gluon. To reconstruct each top quark, we must identify which $b$-jet arises from each top quark decay, and identify the jet pair which arises from each subsequent $W$ boson decay. Techniques developed and validated in the all-hadronic $t\bar{t}$ channel can then be generalized to other collider processes with more complex final states.

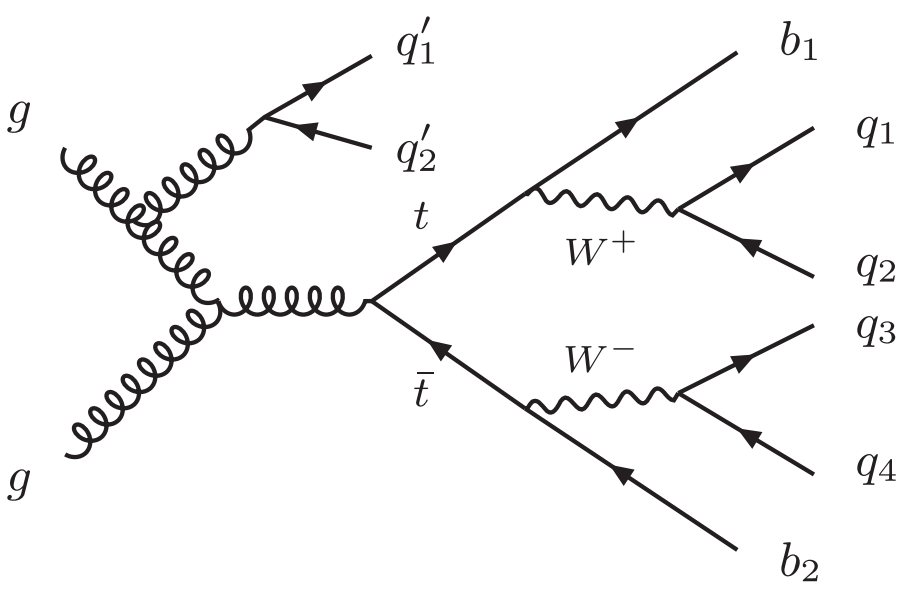

FIG. 1. A next-to-leading-order QCD Feynman diagram for the production of a $t\bar{t}$ pair through the fusion of two gluons, where both $W$ bosons decay hadronically. Two additional quarks ($q'_1$ and $q'_2$) are produced due to initial-state radiation. This leads to a final state characterized by eight jets.

## II. STATE OF THE ART

Kinematic reconstruction algorithms have seen a range of advancements in recent years. Several state-of-the-art machine learning (ML) methods have been developed to reconstruct short-lived particles from detected final states in top quark processes, including SPANet [2,3], Topograph [4], and the Covariant Particle Transformer (*CPT*) [5,6]. The first two techniques seek to identify combinations of final-state jets that correspond to top quark decays. SPANet employs a transformer-based architecture, introducing an attention mechanism that exploits the symmetry properties of the $W$ boson decay to assign the correct final-state jets to their parent particles. Topograph represents events as graphs containing final-state objects, and introduces parent particles (top quarks and $W$ bosons) as additional nodes, with edge connections representing possible decays. Topograph performs traditional edge classification to identify the most probable decay chain and thus identify the correct daughter jets. In contrast, the *CPT* does not classify combinations of jets, instead employing a transformer to directly regress the kinematics of the top quarks from the final-state objects. The use of advanced multivariate techniques in all cases allows these models to explore the complex relations between final states and the kinematics of the parent top quarks, leading to enhanced performance over existing $\chi^2$-minimization-based techniques [2–4,7].

In ML research, extensions of graph representation learning to more powerful and abstract geometric structures have yielded models with greater expressive power, and which are more computationally efficient [8,9]. We hypothesize that leveraging such geometric learning techniques can lead to more effective representations of collider events, and so improve the state-of-the-art in kinematic reconstruction. In this paper, we present *Hypergraph for Particle Event Reconstruction* (HyPER), a machine learning model that uses hypergraph representation learning for particle reconstruction. A hypergraph is a powerful generalization of the traditional graph structure: one which can efficiently represent correlations between an arbitrary number of objects. HyPER builds a hypergraph representation of collider events on top of a traditional *message-passing* graph neural network. The hypergraph structure is then utilized to study higher-order relational information between multiple final-state jets, by constructing and then classifying hyperedges which represent the short-lived particles we wish to reconstruct.

HyPER aims to improve reconstruction performance over current ML methods with a fraction of the model size, while being flexible enough to reconstruct parent particles with an arbitrary number of decay products. The performance of HyPER will be demonstrated in the all-hadronic $t\bar{t}$ channel through comparison with the SPANet and Topograph techniques.[1] The study serves as validation of the HyPER method, and application of HyPER to other collider processes is left as future work.

To the authors' knowledge, this is the first application of hypergraph learning to the reconstruction of resonances in high-energy particle collisions. Complementary research has shown that hypergraph learning can outperform other multivariate methods in reconstructing jets from low-level detector information [10]. Hypergraph learning has also been successfully applied in phenomenological studies of jet substructure [11].

## III. GRAPH AND HYPERGRAPH REPRESENTATIONS

HyPER initially represents each individual collider event as a fully connected digraph,[2] $G = (V, E)$, where $V$ and $E$ are sets consisting of *nodes* and *edges*, respectively. Each final state is interpreted as a node ($i \in V$), and the linkage between any pair of final states as an edge: $E = \{\{i, j\} | i, j \in V \text{ and } i \neq j\}$. HyPER embeds the following kinematic information in the graph structure:

$$\begin{aligned} \mathbf{x}_i^{(0)} &= (p_{Ti}, \eta_i, \phi_i, E_i, b\text{-tag}_i), \\ \mathbf{e}_{j\to i}^{(0)} = \mathbf{e}_{ij}^{(0)} &= (\Delta\eta_{ij}, \Delta\phi_{ij}, \Delta R_{ij}, M_{ij}), \\ \mathbf{u}^{(0)} &= (N_{\text{jets}}, N_{b\text{-tagged}}), \end{aligned} \tag{2}$$

where $\mathbf{x}_i^{(0)}$ and $\mathbf{e}_{ij}^{(0)}$ are the input node and edge attribute vectors, respectively. The definitions of each parameter are given in Appendix A 1. The notation $j \to i$ represents the direction of the edge, thus $\mathbf{e}_{ij}^{(0)} \neq \mathbf{e}_{ji}^{(0)}$. Event-level information, which is not naturally captured in node or edge features, is included in an additional *global* embedding, $\mathbf{u}$.

[1]The *CPT* study [5] is performed on truth-level simulation which does not contain modeling of detector effects on jet kinematics. There is therefore no benchmark for how the *CPT* model will perform on detector-level data, and so we choose not to include it in our study.

[2]A graph whose edges are directional.

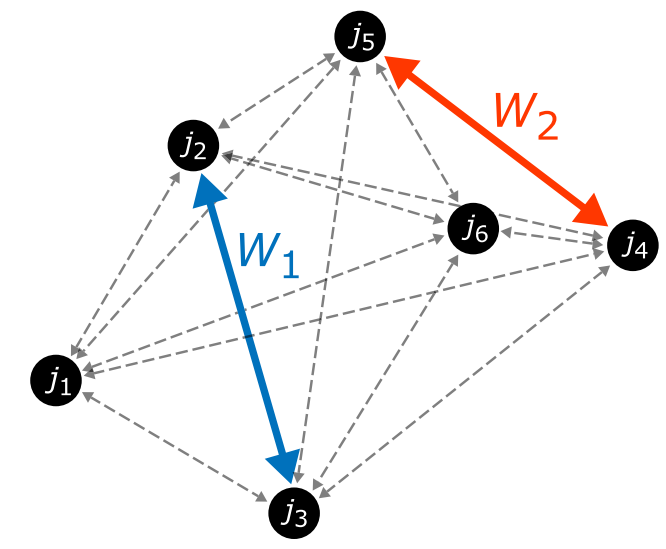

(a) Graph representation

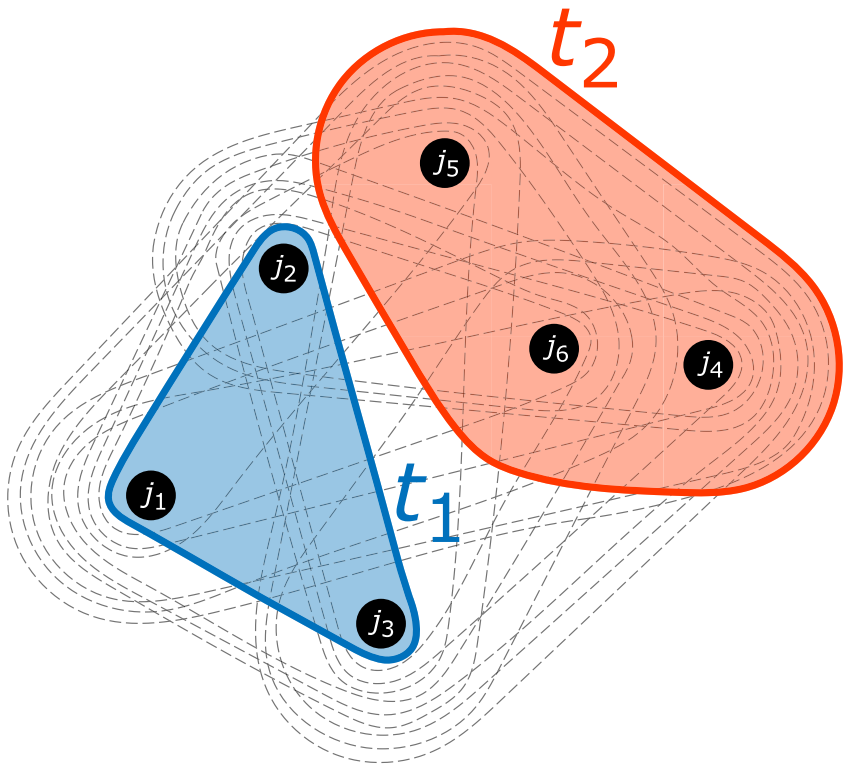

(b) Hypergraph representation

FIG. 2. Illustrations of an event in (a) graph and (b) hypergraph representations. Edges and hyperedges shown in gray dashed lines are the false particle candidates in the reconstruction. The colored edges and hyperedges portray the true parent particles of nodes that they connect.

The graph representation of an event with six final-state jets is shown in Fig. 2(a). In this representation, the decay products of each $W$ boson constitute two distinct nodes: accounting for the direction of the edges, there exist precisely two edges in $E$ representing each true $W$ boson candidate. The edges representing each $W$ boson can be learned using the message-passing technique, which is introduced further in Sec. IV, and whose specific implementation in HyPER is described in Appendix A 2. The task of reconstructing $W$ bosons—and indeed any two-body decay—is an edge classification task. The correct edge candidates are shown by the colored arrows in Fig. 2(a).

In our graph representation, three nodes correspond to the decay products of each top quark. A key limitation of conventional graphs is that edges only encode information between pairs of nodes; top quarks are therefore not effectively represented by traditional edge structures. To address this challenge and ultimately reconstruct top quark pairs, HyPER introduces a set of generalized edges called hyperedges, $\mathcal{E} = \{\tilde{E}_1, \tilde{E}_2, \ldots, \tilde{E}_M\}$, which are sets containing arbitrary numbers of nodes. Specifying identical nodes to the digraph representation, the corresponding hypergraph is defined as $H = (V, \mathcal{E})$.

Top quark candidates can be represented by hyperedges comprised of exactly three nodes:

$$\tilde{E}_m \in \{\{i, j, k\} | i, j, k \in V \quad \text{and} \quad i \neq j \neq k\}, \tag{3}$$

where $m = 1, 2, \ldots, M$, and $M$ is the number of hyperedges.[3] Utilizing the hidden relational information between pairs of nodes learned during the message passing, HyPER constructs hyperedge embeddings, $\tilde{\mathbf{e}}_m$, and uses these embeddings to subsequently classify the hyperedges representing the top quarks. We call the combined use of message passing on graphs with hyperedge classification *blended graph-hypergraph representation learning*.

For an event with a jet multiplicity of six, there are 20 combinations of three jets, leading in principle to 20 candidate top quark hyperedges. This structure is shown in Fig. 2(b), where the dashed gray lines represent all possible three-node hyperedges. The classified $t_1$ and $t_2$ candidates are given by the blue- and orange-shaded hyperedges, respectively. The number of candidate hyperedges could be reduced by imposing certain restrictions, for example requiring that each candidate hyperedge contain a node corresponding to a $b$-tagged jet. We choose to implement HyPER without posing any such requirements on the hyperedge constituents, avoiding any dependence on the accuracy of $b$-tagging algorithms and hence demonstrating the power of the hypergraph method.

## IV. THE HyPER NETWORK

We implement the specific hypergraph representation learning approach outlined in Sec. III through HyPER, whose basic architecture is sketched in Fig. 3. HyPER embeds the kinematic information of an event's final state in the conventional graph structure specified in Eq. (2). These feature-embedded graphs serve as inputs to a message-passing neural network comprised of multiple layers of message-passing aggregation operations. Message passing explores the graph feature space by exchanging information between neighboring nodes, with each message-passing layer leveraging relational information encoded in the graph structure to update the node, edge, and global feature vectors. The message-passing architecture is detailed in Appendix A 2. Each message-passing layer is enumerated with index $t$, with $t = 0$ indicating the initial inputs prior to the first message-passing phase. After the final message-passing operation, $t = T$, the updated attribute vectors $\mathbf{u}^{(T)}, \mathbf{x}_i^{(T)} \in \mathbb{R}^S$ ($\forall \;\; i \in V$) are of dimension $S$; they are preserved and carried on to the top quark reconstruction stage, described in Sec. IV A. The final edge attributes $e_{ij}^{(T)} \in \mathbb{R}$ are used to reconstruct $W$ bosons as outlined in Sec. IV B.

[3] $M$ is given by $\binom{|V|}{3}$ where $|V|$ is the order of the graph.

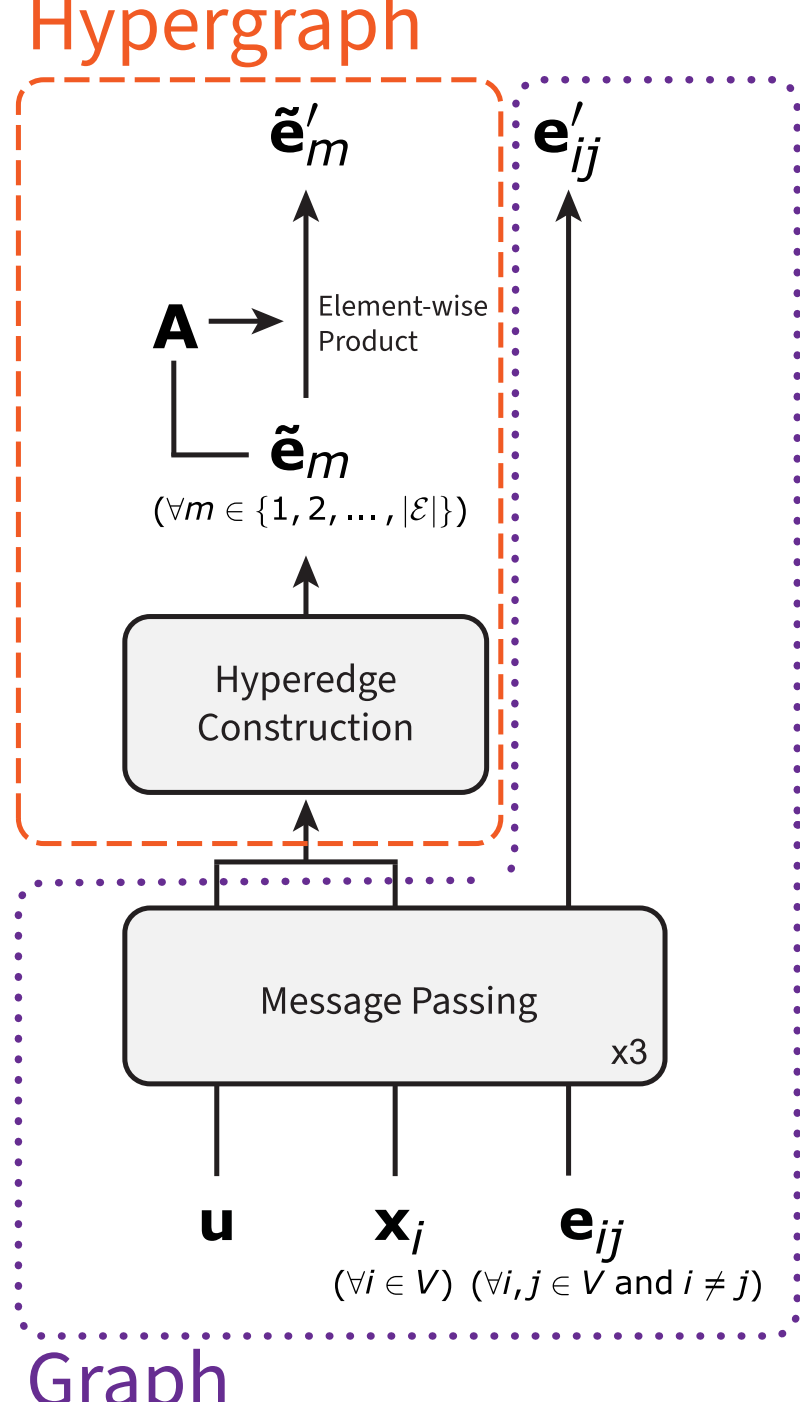

FIG. 3. Network architecture of HyPER. HyPER uses a mixed graph-hypergraph representation. The network components that operate on the graph are enclosed within the purple dotted line, and the components that operate on the hypergraph are enclosed within the orange dashed line.

### A. Hypergraph: Reconstruction of the top quarks

Using a set of predefined hyperedges $\mathcal{E}$ defined according to Eq. (3), HyPER constructs hyperedge features by incorporating node and global attributes inherited from the message passing using an aggregation operator:

$$\tilde{\mathbf{e}}_m = \bigoplus_{l \in \tilde{E}_m} (\mathrm{MLP}(\mathbf{x}_l, \mathbf{u})), \quad \forall\ \tilde{E}_m \in \mathcal{E}, \tag{4}$$

where $l$ is the index of a node connected by the hyperedge $m$, and MLP is a multilayer perceptron. Equation (4) maps the incoming features to a $D$-dimensional hyperedge embedding: $\tilde{\mathbf{e}}_m \in \mathbb{R}^D$.

We seek to assign each hyperedge a score representing the probability that it corresponds to a true top quark. This is achieved by processing each hyperedge embedding with a hyperedge layer that captures the most salient features while projecting the embedding down to a one-dimensional score. Inspired by dynamic graph attention mechanisms [12], the hyperedge layer is defined as

$$\tilde{e}'_m = \mathrm{Sigmoid}(\mathrm{MLP}[\mathbf{A}_m \circ \mathrm{ReLU}(\tilde{e}_m \mathbf{W}^{\mathrm{T}}), \tilde{e}_m]), \tag{5}$$

where $\mathbf{W} \in \mathbb{R}^{D \times D}$ is a learnable weight matrix that linearly transforms hyperedge features. Nonlinearity is introduced to the transformed features using a rectified linear unit [13] (ReLU). For each hyperedge we derive a unique vector $\mathbf{A}_m = (A_{m1}, A_{m2}, \dots, A_{mD})$ from the incoming features of all hyperedges, with each element given by

$$A_{md} = \frac{\exp(\tilde{e}_{md})}{\sum_{m=0}^{M} \exp(\tilde{e}_{md})}, \tag{6}$$

where $A_{md} \in (0, 1)$ represents the importance of a feature $d \in \{1, 2, \dots, D\}$ to a given top quark candidate $m$. Weighted hyperedge attributes are produced by applying the element-wise product (denoted by $\circ$ in Eq. (5)) between derived vectors $\mathbf{A}_m$ and the transformed hyperedge features. Together with the unweighted hyperedge features, they are processed by an MLP for dimensionality reduction, and transformed with the sigmoid function to produce a soft probability: $\tilde{e}'_m \in (0, 1)$. It represents the probability that hyperedge $\tilde{E}_m$ contains the true top quark decay products. Top quark pairs are reconstructed by selecting the two independent, highest-scoring hyperedges.

### B. Graph: Identification of the $W$ boson

The graph edge attributes obtained from the message passing, $e_{ij}^{(T)}$, are transformed with the sigmoid function, yielding a probability that the edges correspond to the actual $W$ boson. Since the ordering of the decays is invariant in the context of reconstruction, edges with the same end points are combined using a permutation invariant function:

$$e'_{ij} = \mathrm{mean}(e_{j \to i}, e_{i \to j}), \tag{7}$$

thus, $e'_{ij} = e'_{ji}$. The reconstruction of $W$ bosons then follows the reconstruction of the top quarks: if $\tilde{E}_t$ is the hyperedge representing the selected top quark, all permissible $W$ boson candidates exist as edges whose nodes are contained in $\tilde{E}_t$. The highest-scoring edge which meets this requirement,

$$\max(e'_{ij}), \qquad i, j \in \tilde{E}_t, \quad i \neq j, \tag{8}$$

is the reconstructed $W$ boson.

### C. Training

The graph and hypergraph components of HyPER are combined through the introduction of a per-graph loss function:

$$\mathcal{L} = \alpha \mathcal{L}_{\mathrm{BCE}}(\mathcal{E}, \hat{\mathcal{E}}) + (1 - \alpha) \mathcal{L}_{\mathrm{BCE}}(E, \hat{E}), \tag{9}$$

where $\mathcal{L}_{\mathrm{BCE}}(\mathcal{E}, \hat{\mathcal{E}})$ is understood to be the mean of the binary cross-entropy (BCE) loss across all hyperedge scores in the hypergraph, $\mathcal{L}_{\mathrm{BCE}}(E, \hat{E})$ the equivalent for graph edge scores, and $\hat{\mathcal{E}}$ and $\hat{E}$ the hyperedge and edge truth labels, respectively. The hyperparameter $\alpha \in [0, 1]$ is introduced to adjust the training importance of graph and hypergraph components. Batching, where multiple graphs are aggregated to improve computational efficiency, is employed. HyPER uses a batch size of 4096. To maximize the event reconstruction efficiency, we compute the loss for each event; the loss of each batch is then the average loss across all events within that batch.

The network is trained with the Adam optimizer [14]. Tuning is performed, with the details of the tuning procedure and a summary of the chosen hyperparameters described in Appendix A 4. Specifically, the message-passing embedding dimension is set to $S = 64$, the dimension of the hyperedge embedding to $D = 128$, and the parameter $\alpha$ to 0.8. Individual tuning is recommended for each individual HyPER use case.

## V. SIMULATION

A total of $6 \times 10^7$ $t\bar{t}$ events are simulated at a center of mass energy of $\sqrt{s} = 13$ TeV. Their matrix elements (MEs) are generated using MADGRAPH5_aMC@NLO (v2.9.16) [15] at next-to-leading order in QCD, interfaced to the five-flavor scheme NNPDF3.0NLO parton distribution function set [16]. The top quark mass is set to $m_t = 172.5$ GeV. We consider only hadronic decays of the top quarks, and these decays are modeled using MADSPIN [17]. The factorization ($\mu_F$) and renormalization scales ($\mu_R$) are set to $\mu_F = \mu_R = m_T$, where $m_T = \sqrt{m_t^2 + 0.5 \times (p_T^2(t) + p_T^2(\bar{t}))}$ is the transverse mass of the top quark. Events are matched to PYTHIA8 (v8.306) [18] to simulate parton shower evolution, hadronization, and multiparton interactions. The simulation of the detector response is performed by DELPHES (v3.5.0) [19] with a parametrization corresponding to the ATLAS detector [20]. Jets are defined using the anti-$k_t$ [21] jet-finding algorithm, as implemented in FASTJET (v3.4.0) [22], with the cone radius parameter set to $R = 0.4$. We impose a selection which mimics that which would be imposed on real data: reconstructed jets must have a minimum $p_T$ of 25 GeV and an absolute pseudorapidity within the ATLAS detector limit of $|\eta| < 2.5$. Jets containing $b$-hadrons ($b$-jets) are tagged using a $p_T$-dependent tagging efficiency (misidentification rate) based on [23].

Supervised learning requires a training sample where the desired outputs are labeled. In the context of this study, the true origin of each jet is established using a generic $\Delta R$ matching scheme. The matching procedure first identifies the outgoing hard particles of the MEs using the generator history, then calculates the angular distance $\Delta R = \sqrt{\Delta\eta^2 + \Delta\phi^2}$ between each pair of clustered jets and hard particles. A jet is considered to have originated from a particle if the angular distance between them satisfies $\Delta R < 0.4$. Only the closest jet in a particle's vicinity is considered to avoid multiple jets being matched to the same particle.

Of the events that pass the kinematic selections, 90% are allocated for model training, with 5% used for model validation and 5% for model performance evaluation. We impose preselection requirements corresponding to an all-hadronic $t\bar{t}$ selection that would be imposed on real collider data. This selection requires each generated event to contain at least six jets, of which at least two must be considered $b$-tagged. A particle is considered *identifiable* if all of its decay products have been matched using the $\Delta R$ truth-matching procedure. Events are also required to contain at least one identifiable $W$ boson. After the preselection, 7,921,299 events remain for training, 439,692 for validation, and 440,069 for testing. Imposing the condition that both top quarks should be identifiable reduces the number of events to 2,116,231 for training, 117,487 for validation, and 116,960 for testing. These events are called *fully matched* events, in contrast to *partially matched* events which have one or more missing jet labels. The highest jet multiplicity recorded in a single event is 20. The full dataset is accessible at [24].

## VI. RESULTS

HyPER is benchmarked against SPANet (v2.2) [2,3] and Topograph [4]. The SPANet model hyperparameters are set to match those prescribed in [3]. Topograph is configured using the recommended settings provided in [4]. All three networks are trained using the same training and validation datasets defined in Sec. V. The three networks have different training strategies for dealing with partially matched events, and missing labels could potentially affect the accuracy of the models [25,26]. We present the results of the three networks trained on fully matched events, and provide performance comparison for models trained on partially matched events in Appendix C. Focusing on models trained on fully matched events, we apply each fully converged network model to the testing dataset. To assess the performance of different networks, we compare their reconstruction results with the truth labels. We introduce two metrics to quantify the performance: per-top efficiency, $\varepsilon_t = N_t^{\mathrm{correct}}/N_t^{\mathrm{identifiable}}$, represents the fraction of top quarks successfully reconstructed from all identifiable ones; per-event efficiency, $\varepsilon_{t\bar{t}} = N_{t\bar{t}}^{\mathrm{correct}}/N_{t\bar{t}}^{\mathrm{FM}}$, is the fraction of fully matched events with both top quarks correctly reconstructed out of all fully matched events. A top quark is considered correctly reconstructed when all its final states are correctly assigned, ignoring the permutations of the two jets originating from the $W$ boson. A correctly reconstructed event has both top quarks successfully reconstructed. The two performance metrics are computed for all three networks in different jet multiplicity regions, and presented in Table I. Similarly, the per-$W$ efficiency metric, $\varepsilon_W$, represents the fraction of $W$ bosons successfully reconstructed from all identifiable ones. This metric is presented for the three networks in Table II. We choose to define $\varepsilon_t$ and $\varepsilon_W$ for all identifiable top quarks and $W$ bosons, regardless of whether the event itself is fully matched or partially matched. We provide analogous efficiencies defined only for fully

TABLE I. Reconstruction performance of HyPER, SPANet, and Topograph evaluated by the event efficiency ($\varepsilon_{t\bar{t}}$) and top quark efficiency ($\varepsilon_t$) calculated in different jet multiplicity regions.

| | $\varepsilon_{t\bar{t}}$ | | | $\varepsilon_t$ | | |
|---|---|---|---|---|---|---|
| Jet multiplicity | HyPER | SPANet | Topograph | HyPER | SPANet | Topograph |
| $= 6$ | $0.816 \pm 0.002$ | $0.801 \pm 0.002$ | $0.805 \pm 0.002$ | $0.663 \pm 0.001$ | $0.667 \pm 0.001$ | $0.658 \pm 0.001$ |
| $= 7$ | $0.672 \pm 0.002$ | $0.654 \pm 0.002$ | $0.649 \pm 0.002$ | $0.654 \pm 0.001$ | $0.649 \pm 0.001$ | $0.638 \pm 0.001$ |
| $\geq 8$ | $0.513 \pm 0.003$ | $0.487 \pm 0.003$ | $0.472 \pm 0.003$ | $0.601 \pm 0.001$ | $0.586 \pm 0.001$ | $0.573 \pm 0.001$ |
| Inclusive | $0.670 \pm 0.001$ | $0.651 \pm 0.001$ | $0.645 \pm 0.001$ | $0.644 \pm 0.001$ | $0.640 \pm 0.001$ | $0.629 \pm 0.001$ |

TABLE II. $W$ boson reconstruction efficiencies for HyPER, SPANet, and Topograph models.

| | $\varepsilon_W$ | | |
|---|---|---|---|
| Jet multiplicity | HyPER | SPANet | Topographs |
| $= 6$ | $0.697 \pm 0.001$ | $0.688 \pm 0.001$ | $0.683 \pm 0.001$ |
| $= 7$ | $0.683 \pm 0.001$ | $0.671 \pm 0.001$ | $0.662 \pm 0.001$ |
| $\geq 8$ | $0.628 \pm 0.001$ | $0.608 \pm 0.001$ | $0.594 \pm 0.001$ |
| Inclusive | $0.675 \pm 0.001$ | $0.662 \pm 0.001$ | $0.653 \pm 0.001$ |

matched events in Appendix B, for direct comparison with the efficiencies presented in [3].

All statistical uncertainties are estimated using the standard expression for the statistical error on efficiencies. The observed differences between the three networks are relatively small compared to the size of the testing set, which contains 116,960 fully matched events, 525,014 identifiable top quarks, and 600,564 identifiable $W$ bosons. Consequently, the estimated statistical uncertainties in each region are practically identical for all networks within the limits of rounding precision.

HyPER shows a higher inclusive event reconstruction efficiency ($\varepsilon_{t\bar{t}}$) of 67.0% compared to the 65.1% and 64.5% obtained by SPANet and Topograph, respectively. While all three networks exhibit a reduction in performance as jet multiplicity increases, HyPER delivers a moderately larger $\varepsilon_{t\bar{t}}$ than its counterparts in all jet multiplicity regions. The hypergraph learning approach is therefore seen to deliver improved performance in $t\bar{t}$ reconstruction. Regarding top quark efficiency, $\varepsilon_t$, SPANet's performance slightly exceeds that of HyPER in the lowest jet multiplicity region; HyPER remediates its inclusive $\varepsilon_t$ with better performance in higher multiplicity regions, reconstructing 64.4% of identifiable top quarks in the inclusive region. The reconstruction efficiency of $W$ bosons, as quantified by per-$W$ efficiency $\varepsilon_W$, is also marginally improved by HyPER. The differences in performance between the three networks are less pronounced than previous comparisons of SPANet and Topograph to the historic $\chi^2$ method [2–4]. Nevertheless, with a network size of $3.45 \times 10^3$, HyPER has shown comparable performance when pitted against the SPANet and Topograph techniques, which have network sizes of $1.7 \times 10^7$ and $6.5 \times 10^6$, respectively.

Using the truth jet assignments, each reconstructed top quark or $W$ boson can be tagged with one of the three orthogonal classes: correct, incorrect, or undetermined. They represent cases where the correct three jets are assigned to the top quark (or two jets to the $W$ boson); where the jet assignment is incorrect; and where the true assignment cannot be determined due to missing truth label(s). A candidate that falls into the correct or incorrect category has by definition an associated truth counterpart. A direct comparison can be drawn between the ground truth[4] and the reconstructed invariant mass spectrum of particles which fall into either the correct or incorrect categories. This comparison is shown in Fig. 4. All three networks successfully recover the top quark and $W$ boson invariant mass peaks, generally retaining the shape of the truth kinematics that the networks were trained upon. Candidates classified under the undetermined category have no truth counterpart due to the presence of one or more unmatched jets. As a consequence, undetermined events cannot be used to assess network performance. However, the invariant mass distributions for the set of undetermined events shows agreement with the shapes of the combined correct and incorrect distributions, peaking around the expected top quark and $W$ boson masses.

To investigate network accuracy across the kinematic phase space, the event efficiency, $\varepsilon_{t\bar{t}}$, is calculated in various truth invariant mass bins of the $t\bar{t}$ system, as illustrated in Fig. 5. All three methods exhibit an increase in $\varepsilon_{t\bar{t}}$ as harder top quark pairs are produced. The two reference methods achieve similar inclusive event efficiency by prioritizing either lower or higher $m_{t\bar{t}}$ bins: Topograph provides better efficiencies in lower $m_{t\bar{t}}$ bins, and SPANet overtakes Topograph in the higher $t\bar{t}$ mass regions. HyPER displays the highest event efficiencies across the spectrum, in particular providing better reconstruction around and above the $t\bar{t}$ production threshold, where the majority of $t\bar{t}$ pairs are produced.

HyPER fulfills its goal of achieving state-of-the-art reconstruction performance with far fewer trainable parameters than other reconstruction techniques. We investigate other performance metrics in an attempt to understand how this superior parameter efficiency may yield additional benefits. Larger networks may exhibit degraded performance when trained on smaller datasets, and in some cases may be more susceptible to overtraining [27,28]. This is important as many experimental measurements do not have access to large simulated samples for network training. Examples include heavy Higgs searches which require generation of simulated samples at a large number of different heavy Higgs mass points; in such cases, each sample can have as little as 100,000 generated events. Figure 6 presents the performance of each network, as quantified by $\epsilon_{t\bar{t}}$, for six different sizes of training datasets.[5] All three models generally exhibit reduced reconstruction performance as the training dataset size decreases. HyPER is the best performing reconstruction technique for all considered training dataset sizes, demonstrating robust reconstruction performance across the range of training datasets.

[4]*Ground truth* is the value of $m_t$ obtained using the truth labels.

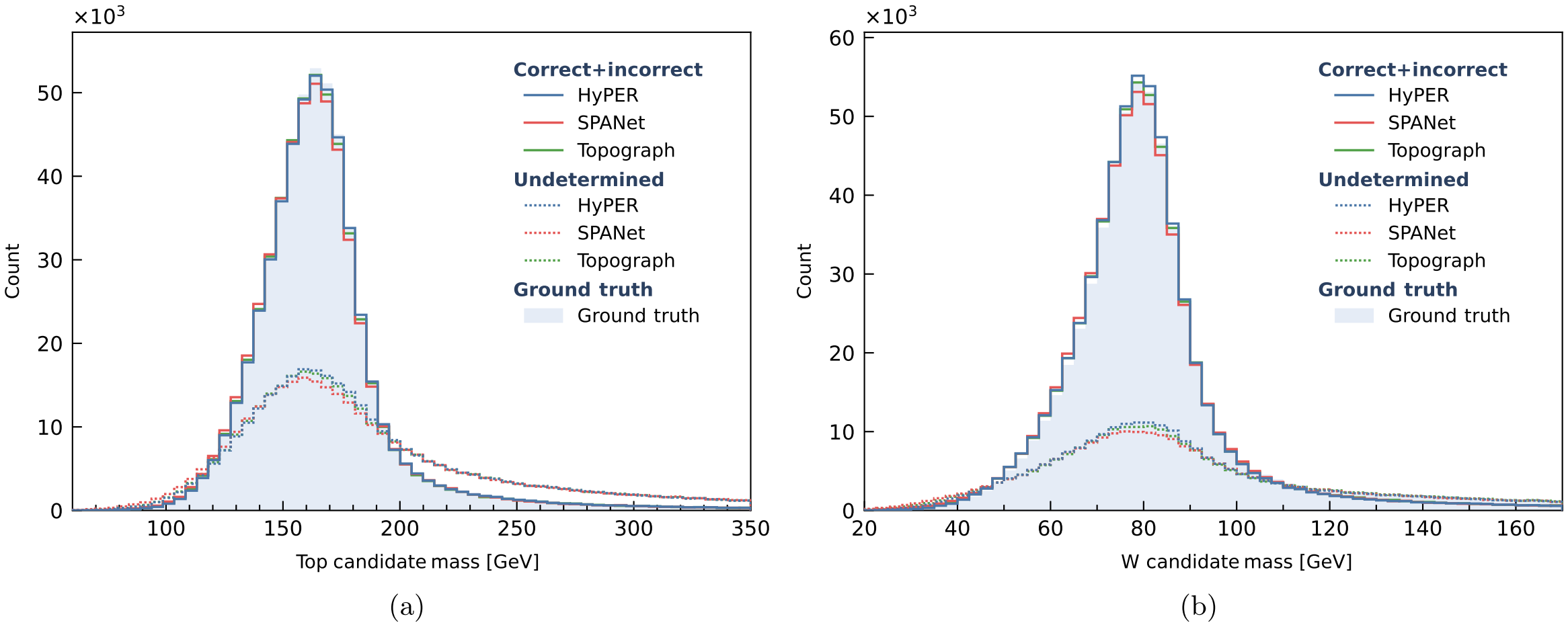

FIG. 4. Invariant mass distributions of reconstructed (a) top quarks and (b) $W$ bosons, shown in combined correct-incorrect (solid) and undetermined (dotted) categories. They are compared to the truth mass distributions (shaded light blue) obtained using jet-assignment labels.

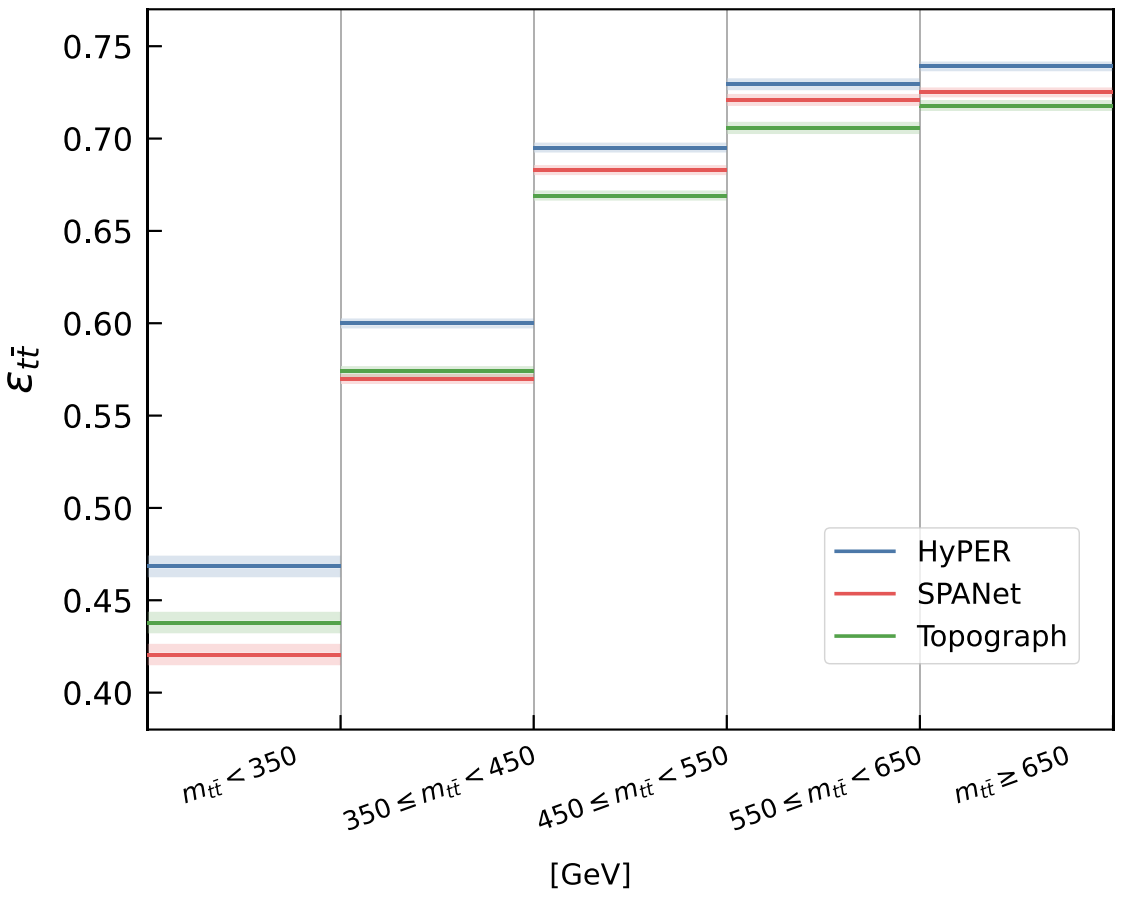

FIG. 5. Event reconstruction efficiencies of HyPER, SPANet, and Topograph, calculated in various truth $t\bar{t}$ invariant mass ($m_{t\bar{t}}$) regions.

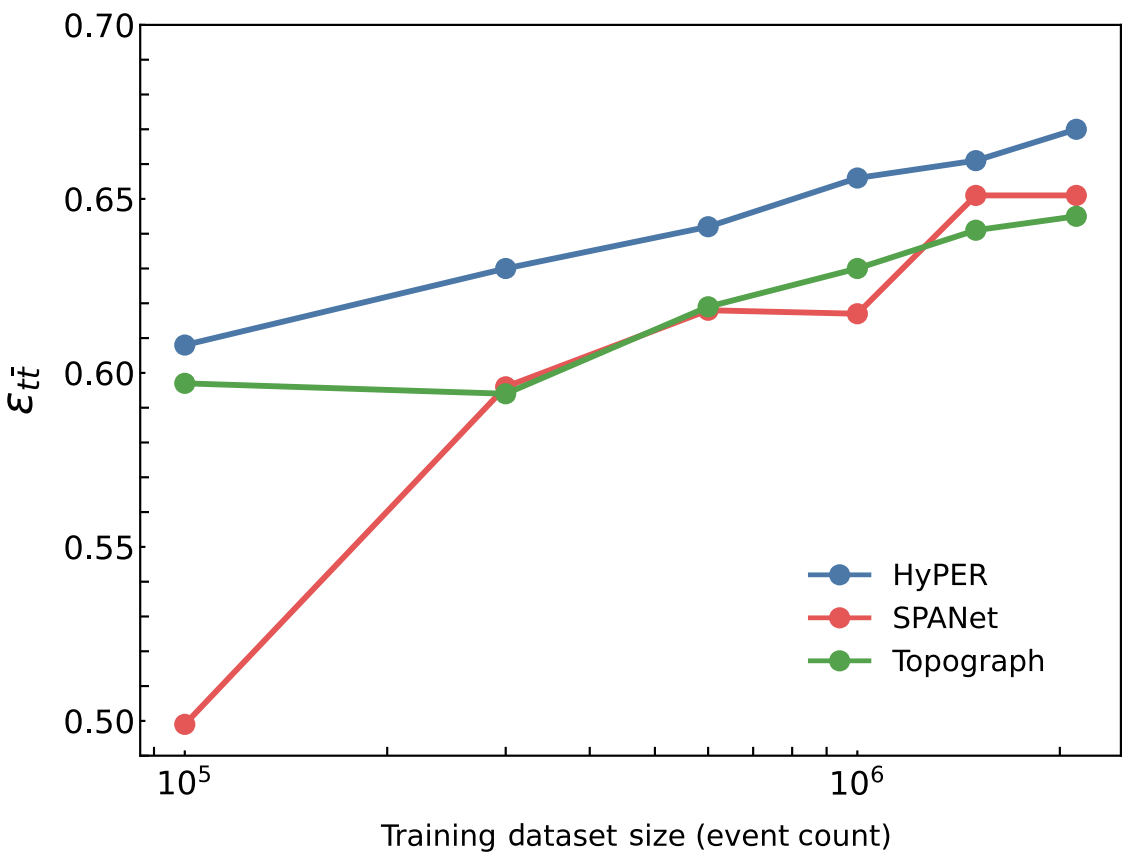

FIG. 6. Comparison of network performance for training datasets of varying sizes. The statistical uncertainty on $\epsilon_{t\bar{t}}$ is identically $\pm 0.001$ for each network for all dataset sizes, as the same testing dataset is used throughout this test.

HyPER contains far fewer trainable parameters than SPANet or Topograph, and this is conjectured to have an impact on computational performance. A metric which is strongly correlated with parameter efficiency is the graphics processing unit (GPU) memory consumed in a typical training epoch. All three networks are trained on one NVidia A100 GPU, and the GPU memory consumption is presented for each network. Batch size will have a large impact on GPU memory consumption; thus for the purposes of this study, the batch size of each network is set to 2048. The GPU memory consumption over a dedicated training of 20 epochs is presented in Fig. 7. HyPER is found to have a far smaller GPU memory footprint; this is attributed to the efficiency of the model in requiring far fewer learnable parameters.

[5]It should be noted that SPANet displayed some instability when trained on datasets containing less than $10^6$ events. In such cases, training was repeated and the highest $\epsilon_{t\bar{t}}$ used for comparison.

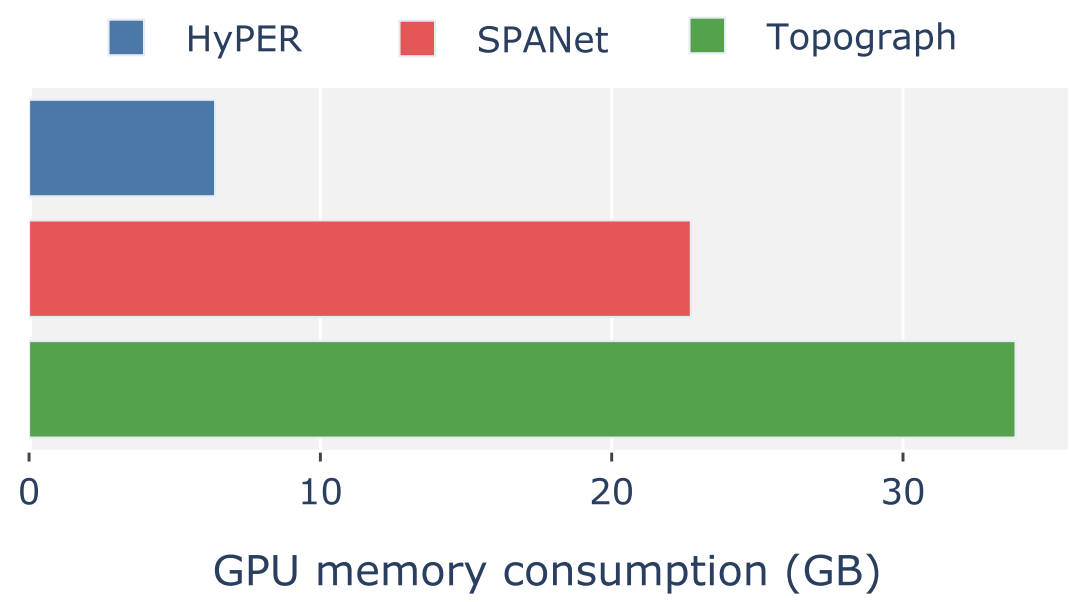

FIG. 7. GPU memory consumption (in gigabytes) for each model, trained across 20 epochs.

## VII. CONCLUSIONS

We have presented HyPER, a novel hypergraph learning methodology for reconstructing short-lived particles in collider events with complex final states. The hypergraph representation learning method was introduced to achieve a more powerful representation of $t\bar{t}$ final states that not only improved reconstruction performance, but did so using a more efficient model with fewer trainable parameters. We presented benchmark tests against two existing state-of-the-art methods, SPANet and Topograph, focusing on reconstructing top quarks and $W$ bosons in the all-hadronic $t\bar{t}$ channel. The HyPER approach yields competitive results, reconstructing 67% of the fully matched events, compared to the 65.1% achieved by SPANet and 64.5% by Topograph, with a network architecture that contains a factor of 20 fewer parameters. HyPER also demonstrated comparatively higher performance in all $t\bar{t}$ invariant mass regions, and remained the most accurate model when trained on a range of training dataset sizes.

HyPER is unique in introducing hypergraph representation learning to high-level event information, a new development in kinematic reconstruction problems. Edge and hyperedge classification were used to reconstruct $W$ bosons and top quarks, respectively, but hyperedges can represent candidate particles which decay into an arbitrary number of final states. In principle, this facilitates the reconstruction of any short-lived SM or beyond the Standard Model particle. The versatility of this approach, applicable to arbitrarily complex collider events, should allow for the effective reconstruction of a wide variety of physics processes. For instance, the associated production of a top quark pair and Higgs boson, $t\bar{t}H$, can be represented using a hypergraph whose structure is identical to that of a $t\bar{t}$ event with an additional target edge label specified for the Higgs boson. Ongoing work also considers the simultaneous reconstruction of final-state neutrino kinematics, with the aim of reconstructing four-top production events in multiple decay channels.

Many experimental measurements require both the reconstruction of heavy states and the classification of the production process, the latter being used to separate a signal from potential backgrounds. These two tasks are often performed by separate multivariate tools, and so the edge and hyperedge outputs of HyPER could be used to construct discriminants for input to downstream classification networks. However, by taking advantage of preexisting message-passing operations, HyPER can in principle be used to perform event classification while simultaneously reconstructing particles. This could be achieved by incorporating an additional loss term that utilizes the output of the global graph attributes, $\mathbf{u}$, whose inherent representation of graphwise information makes it ideal for extracting event-level properties. Despite the introduction of a supplementary classification task, no additional MLPs are introduced; thus HyPER will remain parameter efficient. Further extensions could consider using the full hypergraph representation of events to provide additional discriminatory power between various collider processes, and thus continue the development of hypergraph representation learning in high-level particle physics tasks.

## ACKNOWLEDGMENTS

The authors would like to acknowledge the assistance given by Research IT and the use of the Computational Shared Facility at the University of Manchester. The authors would also like to acknowledge their colleagues in the Manchester HEP group for insightful discussions and support. Y. P. and E. S. are supported by the European Research Council (ERC), under Grant No. 817719. C. B.-S. was supported by the UK Research and Innovation (UKRI), Science and Technology Facilities Council (STFC) under Grant No. ST/P000800/1. B. L and C. B.-S. contributed to this work when they were affiliated with the University of Manchester.

## APPENDIX A: ADDITIONAL INFORMATION ON HyPER

HyPER is a PYTHON project, with code publicly available on GitHub.[6] The HyPER model is created with PYTORCH and PYTORCH-GEOMETRIC libraries [29,30], and the training framework is built using PYTORCH-LIGHTNING [31]. In this section, we describe some of the fundamental components of HyPER, including the network inputs, the mathematical formulation of the message-passing operations, and the general design of the MLPs. We also provide a list of optimized hyperparameters essential for recreating the results presented.

### 1. Graph input kinematics

The parameters $p_{Ti}$ and $E_i$ correspond to the transverse momentum and energy of the $i$-th jet, respectively. Angular terms $\eta_i$ and $\phi_i$ correspond to the pseudorapidity and azimuth angle in the transverse plane, as defined with respect to the standard ATLAS coordinate system. The signed terms $\Delta\eta_{ij}, \Delta\phi_{ij}$ are the corresponding angular separations between the $i$-th and $j$-th jets. The term $\Delta R_{ij}$ represents the Euclidean separation in the $\eta$–$\phi$ plane, as given by $\sqrt{\Delta\eta^2 + \Delta\phi^2}$. The boolean $b$-tag$_i$ labels whether the jet is $b$-tagged. The term $M_{ij}$ is the invariant mass of the $ij$ jet pair. The global terms $N_{\text{jets}}$ and $N_{b\text{-tagged}}$ denote the total number of jets in the event and the number that are $b$-tagged, respectively. Input features $p_{Ti}$, $E_i$ and $M_{ij}$ are log normalized before being passed into the network.

[6] See https://github.com/tzuhanchang/HyPER.git.

### 2. Message passing

Message passing is a convolution operation that has been generalized to operate on non-Euclidean domains, such as graph-structured data [32,33]. The message-passing technique used in HyPER is adopted from [32–35]. A complete message-passing layer consists of three mathematical operations that update the edge, node, and global attributes.

For time step $t$, the message-passing process begins by gathering all available features in the graph and updating the edges through an aggregation operation:

$$\mathbf{e}_{ij}^{(t+1)} = \text{MLP}_{\mathbf{e}}^{(t)}\left(\mathbf{x}_i^{(t)}, \mathbf{x}_j^{(t)}, \mathbf{e}_{ij}^{(t)}, \mathbf{u}^{(t)}\right),$$

where $i$ and $j$ are two adjacent nodes. A *message*, $\mathbf{m}_{ij}$, is then calculated for each updated edge using a message function:

$$\mathbf{m}_{ij}^{(t)} \equiv \mathbf{m}_{j\to i}^{(t)} = \text{MLP}_{\mathbf{m}}^{(t)}\left(\mathbf{x}_i^{(t)}, \mathbf{x}_j^{(t)}, \mathbf{e}_{ij}^{(t+1)}\right).$$

Messages are signed, $\mathbf{m}_{ij} \neq \mathbf{m}_{ji}$: they have identical directions to the edges they are carried by. The notation $j \to i$ indicates the source and recipient nodes of the message, respectively. Once the recipient node $i$ has received all the messages from its direct neighbors, these messages are summarized, updating the recipient node to its next state as follows:

$$\mathbf{x}_i^{(t+1)} = \text{MLP}_{\mathbf{x}}^{(t)}\left[\mathbf{x}_i^{(t)}, \mathbf{u}^{(t)}, \underset{\forall j: j\to i}{\text{mean}}\left(\mathbf{m}_{ij}^{(t)}\right), \underset{\forall j: j\to i}{\max}\left(\mathbf{m}_{ij}^{(t)}\right)\right].$$

The global attribute vector, $\mathbf{u}$, is updated to its next instance by collecting information from updated nodes. This process is formulated as

$$\mathbf{u}^{(t+1)} = \text{MLP}_{\mathbf{u}}^{(t)}\left[\mathbf{u}^{(t)}, \underset{\forall i}{\text{mean}}\left(\mathbf{x}_i^{(t+1)}\right), \underset{\forall i}{\max}\left(\mathbf{x}_i^{(t+1)}\right)\right].$$

This concludes a full message-passing layer. The message-passing operation is repeated, with outputs of each the previous layer forming inputs to each subsequent layer, until the number of maximum iterations $T$ is reached. In this study, we implement three message-passing layers.

TABLE III. A list of hyperparameters used to set up HyPER for the presented study.

| Name | Notation | Value |
|---|---|---|
| Number of message-passing iterations | $T$ | 3 |
| Message dimensionality | $S$ | 64 |
| Hyperedge dimensionality | $D$ | 128 |
| Learning rate | $\gamma$ | 0.001 |
| Loss mixing | $\alpha$ | 0.8 |
| Dropout | $p$ | 0.01 |
| Batch size | $B$ | 4096 |

### 3. MLP design

Multilayer perceptions are the fundamental building blocks of HyPER. The MLPs used in HyPER are designed with the same architecture, varying only in the dimensionalities of the input and output attribute vectors. Each MLP contains two fully connected hidden layers, each formulated as follows:

$$\phi(\mathbf{h}_n, n) = \text{Dropout}[\text{ReLU}(\mathbf{h}_n \mathbf{W}_n^{\text{T}} + \mathbf{b}_n)], \quad \text{(A1)}$$

where $\mathbf{h}_n$ is the input attributes of the hidden layer $n$. The dimensionalities of the weight matrices $\mathbf{W}_n$ are determined by the input and output dimensions of the MLP, and $\mathbf{b}_n$ are learnable biases. The activation function ReLU [13] in Eq. (A1) introduces nonlinearity to the model. Dropout [36] is a regularization technique employed to prevent overfitting.

### 4. Hyperparameters

A set of optimized hyperparameters used in the study is shown in Table III. Hyperparameter tuning is performed by the OPTUNA (v3.4) optimization framework [37]. A total of 100 HyPER models are tested, each with a distinct set of hyperparameters as sampled by OPTUNA. Each model is trained for ten epochs. Pruning, which terminates unpromising trials at early stages, is employed to enhance the efficiency of the tuning process. The training and validation datasets used for tuning are randomly sampled from the primary training dataset; they contain 1 M and 50 K events, respectively.

The learning rate, $\gamma$, is updated during training according to the monitored hyperedge accuracy as evaluated on the validation samples. This is done using a learning rate schedule scheme, which reduces $\gamma$ by a factor of 0.8 if the hyperedge accuracy is not seen to increase in ten epochs. Training is terminated when the hyperedge accuracy does not improve for 50 epochs.

## APPENDIX B: FULLY MATCHED EFFICIENCIES

Per-top efficiency $\varepsilon_t$ and per-$W$ efficiency $\varepsilon_W$ are defined in Sec. VI as the ratio of correctly reconstructed particles to all identifiable particles of that type, regardless of whether the event is fully matched or partially matched. The related metrics $\varepsilon_t^{\mathrm{FM}}$ and $\varepsilon_W^{\mathrm{FM}}$ are the efficiencies defined only for fully matched events. This provides a direct comparison with [3]. The results are tabulated in Table IV. Each metric is larger compared to the efficiencies computed for the combined fully matched and partially matched events. Results are similar for each network in all jet multiplicity regions, with HyPER displaying the highest efficiencies in all categories.

TABLE IV. Performance of HyPER, SPANet, and Topograph in reconstructing individual top quarks and $W$ bosons in fully matched events.

| | $\varepsilon_t^{\mathrm{FM}}$ | | | $\varepsilon_W^{\mathrm{FM}}$ | | |
|---|---|---|---|---|---|---|
| Jet multiplicity | HyPER | SPANet | Topograph | HyPER | SPANet | Topograph |
| $=6$ | $0.853 \pm 0.001$ | $0.838 \pm 0.001$ | $0.843 \pm 0.001$ | $0.893 \pm 0.001$ | $0.881 \pm 0.001$ | $0.884 \pm 0.001$ |
| $=7$ | $0.769 \pm 0.002$ | $0.751 \pm 0.002$ | $0.750 \pm 0.002$ | $0.814 \pm 0.001$ | $0.798 \pm 0.001$ | $0.797 \pm 0.001$ |
| $\geq 8$ | $0.668 \pm 0.003$ | $0.644 \pm 0.002$ | $0.636 \pm 0.002$ | $0.718 \pm 0.002$ | $0.695 \pm 0.002$ | $0.685 \pm 0.002$ |
| Inclusive | $0.765 \pm 0.001$ | $0.746 \pm 0.001$ | $0.745 \pm 0.001$ | $0.810 \pm 0.001$ | $0.793 \pm 0.001$ | $0.791 \pm 0.001$ |

## APPENDIX C: PARTIAL EVENT TRAINING

The $\Delta R$ truth-matching scheme often fails to associate jets with their parent particles, resulting in missing jet origin labels. These events are tagged as partially matched. SPANet can be trained on events with at least one identifiable top quark, and for HyPER and Topograph, at least one identifiable $W$ boson is required. For HyPER, in the case where an event has no identifiable top quarks due to two missing $b$-quark labels, $\alpha$ in Eq. (9) is set to zero, such that only the loss of the $W$ boson is considered. We train all three networks on expanded training and validation datasets (derived from the simulation described in Sec. V) which now include the partially matched events. Network performance is assessed using the per-event efficiency ($\varepsilon_{t\bar{t}}$) and per-top efficiency ($\varepsilon_t$). Results are shown in Table V.

TABLE V. Per-event efficiency ($\varepsilon_{t\bar{t}}$) and per-top efficiency ($\varepsilon_t$) evaluated in various jet multiplicity regions using models trained on partially matched datasets.

| | $\varepsilon_{t\bar{t}}$ | | |
|---|---|---|---|
| Jet multiplicity | HyPER | SPANet | Topograph |
| $=6$ | $0.798 \pm 0.002$ | $0.808 \pm 0.002$ | $0.787 \pm 0.002$ |
| $=7$ | $0.663 \pm 0.002$ | $0.667 \pm 0.002$ | $0.632 \pm 0.002$ |
| $\geq 8$ | $0.513 \pm 0.003$ | $0.503 \pm 0.003$ | $0.462 \pm 0.003$ |
| Inclusive | $0.661 \pm 0.001$ | $0.663 \pm 0.001$ | $0.631 \pm 0.001$ |

| | $\varepsilon_t$ | | |
|---|---|---|---|
| Jet multiplicity | HyPER | SPANet | Topograph |
| $=6$ | $0.709 \pm 0.001$ | $0.698 \pm 0.001$ | $0.677 \pm 0.001$ |
| $=7$ | $0.675 \pm 0.001$ | $0.668 \pm 0.001$ | $0.644 \pm 0.001$ |
| $\geq 8$ | $0.612 \pm 0.001$ | $0.601 \pm 0.001$ | $0.570 \pm 0.001$ |
| Inclusive | $0.672 \pm 0.001$ | $0.663 \pm 0.001$ | $0.638 \pm 0.001$ |

All three models exhibit increased $\varepsilon_t$ compared to models trained with fully matched events. This is due to the significantly expanded training set, which allows the networks to access more identifiable top quarks during training. HyPER is the best performing model in this category. SPANet's per-event reconstruction is slightly improved when training on partially matched events, while HyPER and Topograph display minor reductions in performance. The configuration with the best per-event performance remains the HyPER model trained on fully matched events, as can be seen through a comparison of Tables I and V. Though not tested here, HyPER's performance could potentially be improved by scaling up the dimensionalities of message and hyperedge attribute vectors for training on partially matched events.

## APPENDIX D: HyPER OUTPUTS

The reconstruction procedure selects two independent highest-scoring hyperedges to form top quark candidates. An enclosed highest-scoring graph edge is chosen for each selected hyperedge as its corresponding $W$ candidate. We record the scores for the reconstructed top quarks and $W$ bosons, denoted by $\tilde{e}'_{\mathrm{top}}$ and $e'_W$, respectively. The two distributions are shown in Figs. 8(a) and 8(b), subdivided into "correct," "incorrect," and "undetermined" categories as before. The scores $\tilde{e}'_{\mathrm{top}}$ and $e'_W$ demonstrate reasonable separation between the three categories.

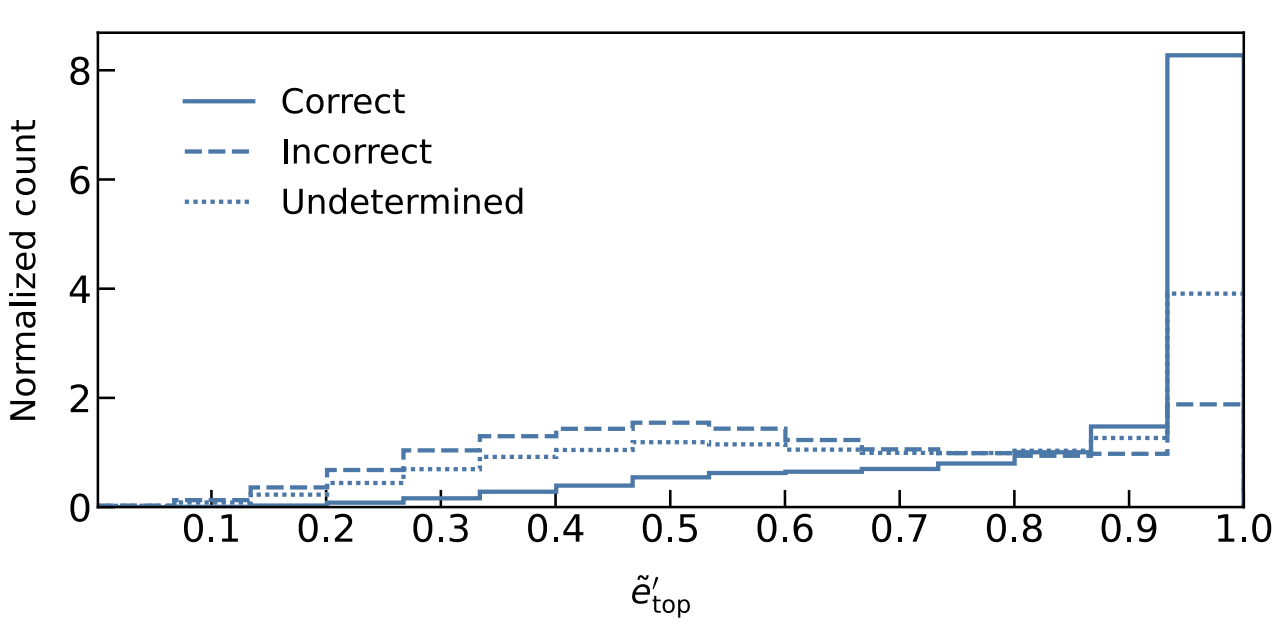

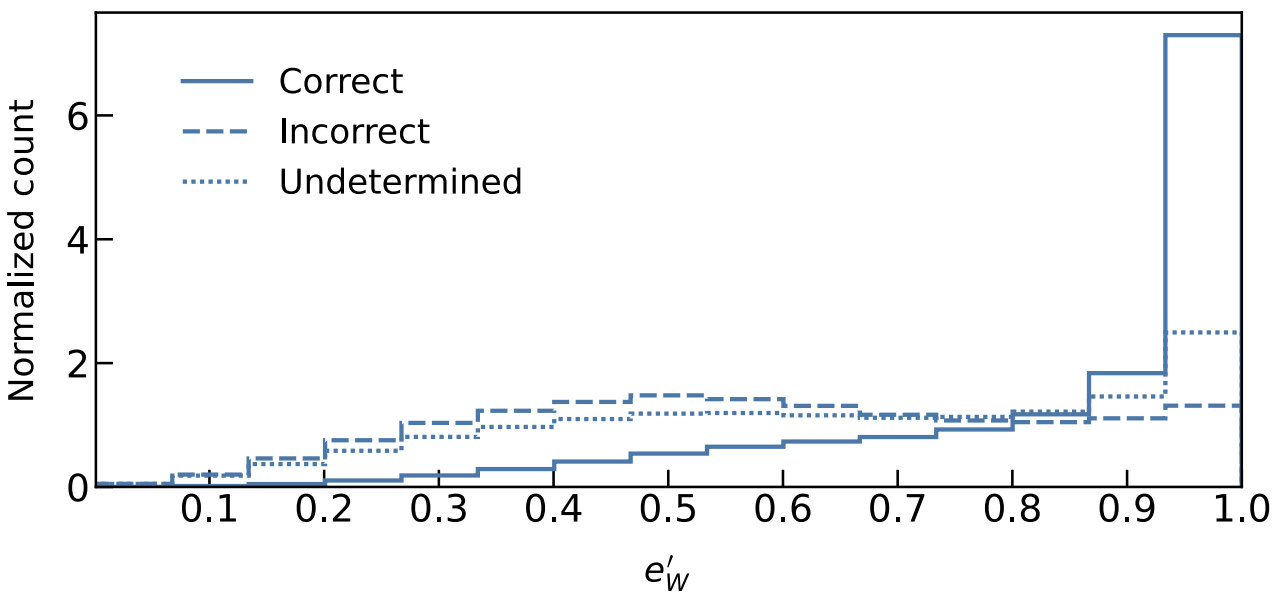

FIG. 8. HyPER output distributions of reconstructed (top panel) top quarks, $\tilde{e}'_{\mathrm{top}}$, and (bottom panel) $W$ bosons, $e'_W$.

We propose that these scores could be used as discriminants for enhancing the efficiency of correctly reconstructed particles. Though not investigated in this paper, one could impose a cut on $\tilde{e}'_{\mathrm{top}}$ and $e'_W$ to increase the confidence that the reconstructed objects do correspond to the true particles. For instance, one could require that both top quarks and $W$ bosons in an event must pass the selection requirements $\tilde{e}'_{\mathrm{top}} > 0.8$ and $e'_W > 0.8$. This region constitutes 46.1% of all fully matched events and has an inclusive event efficiency of $\varepsilon_{t\bar{t}} = 0.923$. The optimal performance and statistics can be achieved by fine-tuning these confidence requirements.

## APPENDIX E: SUPPLEMENTARY PLOTS

We present two additional visualizations of the networks' reconstruction performance. Figure 9 shows per-jet assignment efficiencies using a separate confusion matrix for each network. Figure 10 shows the per-event and per-top efficiencies for each network as a function of jet multiplicity.

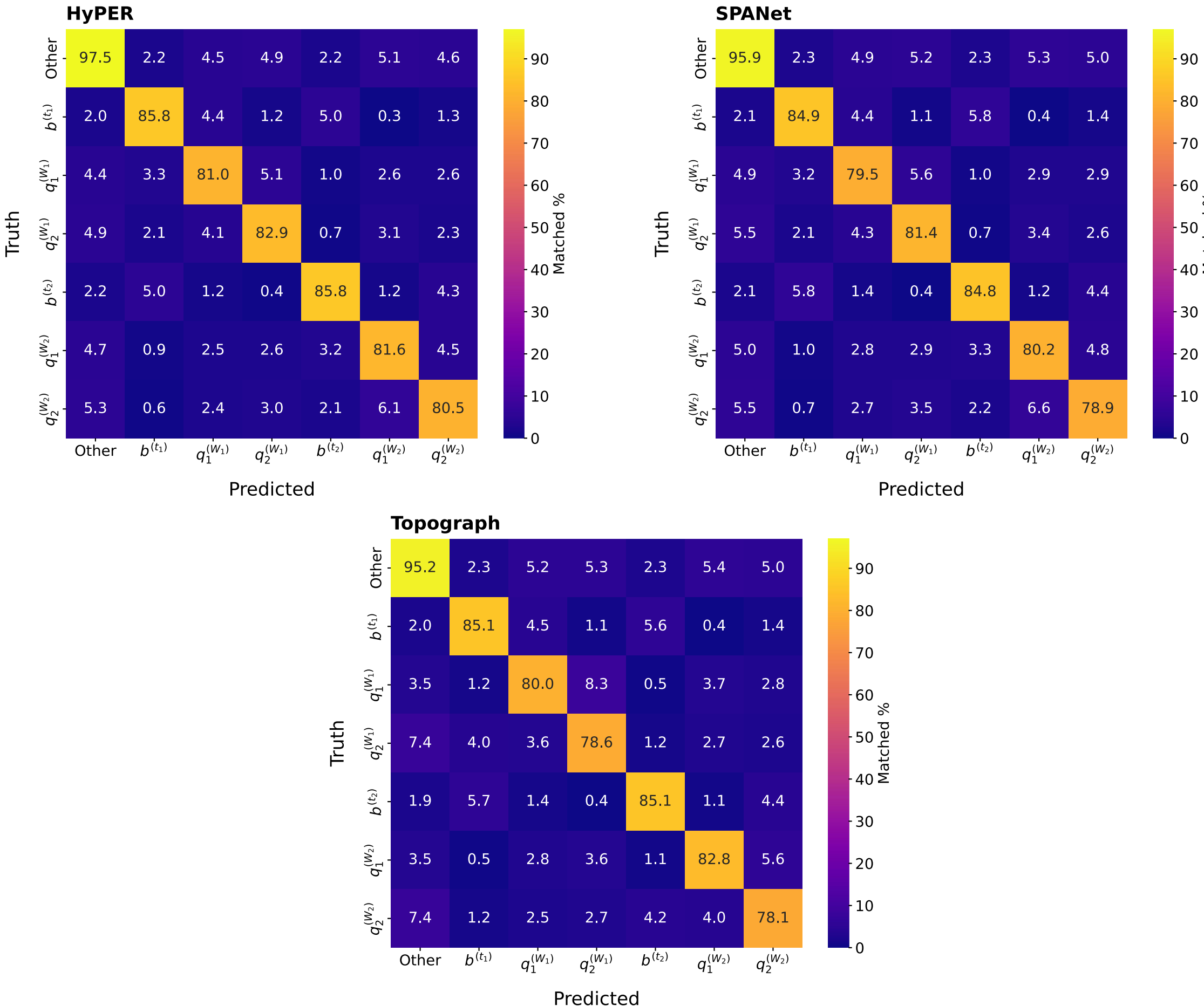

FIG. 9. Per-jet assignment efficiencies of the reconstruction networks visualized using normalized confusion matrices. Each row represents the networks' predictions of the jet origins, and each column represents the true origins of the jets. The category "other" corresponds to jets which do not originate from the decay of one of the top quarks. This category sums to greater than unity as more than one additional jet can exist in any event. The matrices are evaluated using the fully matched events. HyPER consistently displays the highest percentage of correctly matched jets.

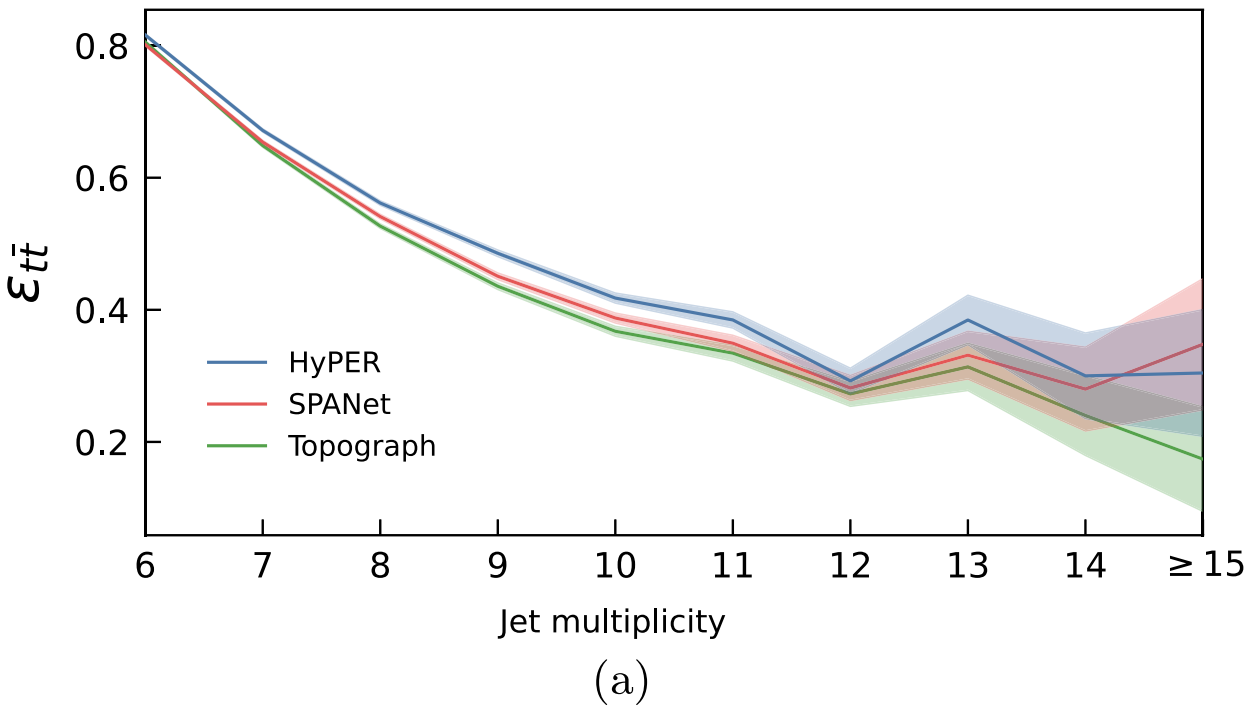

(a)

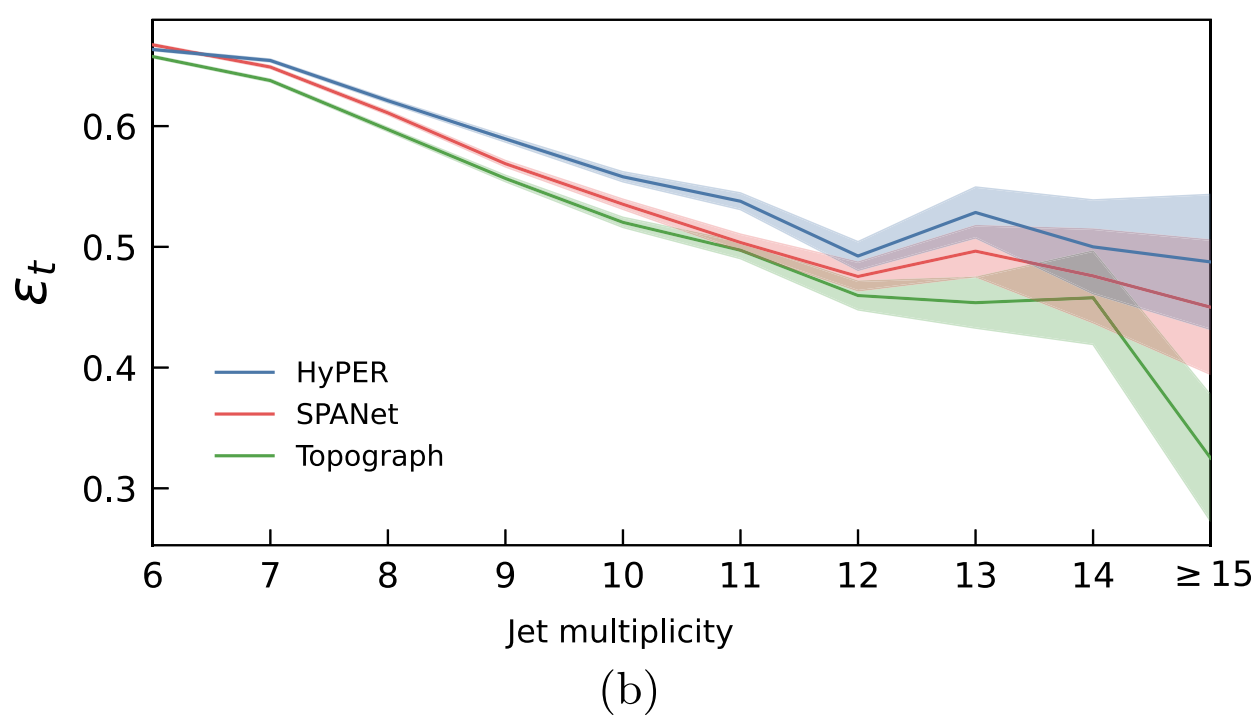

(b)

FIG. 10. Comparison of (a) per-event and (b) per-top efficiencies for HyPER, SPANet, and Topograph in various jet multiplicity regions.